# Compact Stirling cooling of astronomical detectors

Gert Raskin[1], Johan Morren[1], Wim Pessemier[1], Jesus Perez Padilla[2] and Jeroen Vandersteen[3]

[1]*Instituut voor Sterrenkunde, KU Leuven University, Belgium;* [2]*Mercator telescope, La Palma, Spain;* [3]*RHEA - European Space Agency, ESTEC, The Netherlands*

**ABSTRACT**

*MAIA, a three-channel imager targeting fast cadence photometry, was recently installed on the Mercator telescope (La Palma, Spain). This instrument observes a 9.4 x 14.1 arcmin field of view simultaneously in three color bands (u, g and r), using three of the largest (un-) available frame-transfer CCDs, namely the 2k x 6k CCD42-C0 from e2v. As these detectors are housed in three separate cryostats, compact cooling devices are required that offer sufficient power to cool the large chips to a temperature of 165K. We explored a broad spectrum of cooling options and technologies to cool the MAIA detectors. Finally, compact free-piston Stirling coolers were selected, namely the CryoTel MT cryo-coolers from SUNPOWER, that can extract 5W of heat at a temperature of 77K. In this contribution we give details of the MAIA detector cooling solution. We also discuss the general usability of this type of closed-cycle cryo-coolers for astronomical detectors. They offer distinct advantages but the vibrations caused by mechanical cryo-cooling impose a serious drawback. We present a solution to mechanically decouple the coolers from the instrument and show some results of how this reduces the vibrations to a level that is acceptable for most applications on astronomical telescopes.*

## 1. INTRODUCTION

Most optical astronomical instruments rely on a CCD as detector. The performance of the CCD depends, among others, on its temperature. Even in the absence of light, thermal noise in the silicon of the CCD will create free electrons or dark current that will be collected as signal in the pixels of the detector. These dark electrons become an indistinguishable part of the signal and add to both total flux and noise. Dark current in silicon is a strong function of temperature, as it will approximately double for each temperature increase of 6K. This implies that both dark current and dark current shot noise can be suppressed or even completely eliminated by reducing the temperature of the detector.

The three CCDs that are used in the MAIA imager also require cooling in order to limit their dark current. As these CCDs are housed in three separate cryostats, three individual coolers are required. Therefore, we need a cooling system that is compact and efficient. Various technologies exist to cool a CCD detector. In this contribution, we discuss the detector cooling system that we developed for the MAIA imager and that makes use of compact Stirling coolers. We show that this can be a convenient and attractive solution for many other scientific imaging instruments.

---

[1] gert.raskin@ster.kuleuven.be

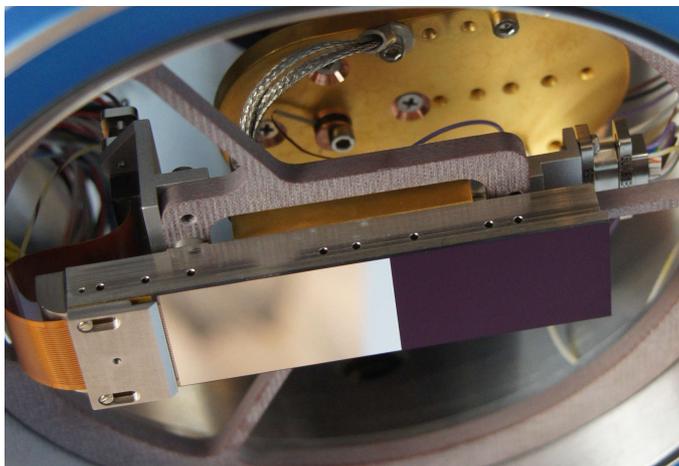

*Figure 1 Picture of a CCD42-C0 frame-transfer detector mounted in its cryostat.*

## 2. THE MAIA INSTRUMENT

MAIA, an acronym for Mercator Advanced Imager for Asteroseismology, is a three-channel instrument that targets fast-cadence three-color photometry [1]. This instrument observes a 9.4 x 14.1 arcmin$^2$ field of view simultaneously in the *u*, *g* and *r* wavelength bands on three CCD42-C0 detectors. These are very large frame-transfer detectors (2k x 6k 13.5 micron pixels) that were originally developed and built by e2v (UK) for ESA's Eddington space mission. In figure 1, we show a picture of such a detector mounted on a fiberglass spider in the *u*-channel cryostat. After ESA cancelled the Eddington mission, the Institute of Astronomy of the KU Leuven University (Belgium) received these detectors on permanent loan to build the MAIA instrument, provided that they would be used for one of the original science goals of the Eddington mission. MAIA targets building up time series of three-color photometric observations for asteroseismology research, with specific emphasis on sub-dwarf and white dwarf single and binary stars. In 2012, MAIA was installed on the 1.2-m Mercator telescope [2] at the Roque de los Muchachos Observatory on La Palma (Canary Islands, Spain). Instrument commissioning took place in 2013. A picture of MAIA mounted at the Nasmyth focus of the telescope and a sketch of the optical layout of the instrument are shown in figure 2.

## 3. MAIA DETECTOR COOLING

### 3.1 Detector temperature requirement

Although the most important science case for MAIA mainly needs short integration times, instrument versatility also required the possibility of obtaining long exposures. Even when we observe the faintest targets, the detector dark current should still be a negligible noise source. This means that it should be significantly smaller than the flux generated by the new-moon sky. For MAIA at the Mercator telescope, this flux can be as small as 0.1 e⁻/pixel/s in the *u* band. Hence, the dark current should not exceed 0.05 e⁻/pixel/s. According to laboratory measurements, the CCD42-C0 detectors reach this dark current level at temperatures below 190K. Alternatively, the dark current can already be reduced at a substantially higher temperature by operating the detector in inverted state. However, this inversion also reduces the full-well capacity and this limits the dynamic range of the detector. Therefore, the solution of inverted operation was discarded.

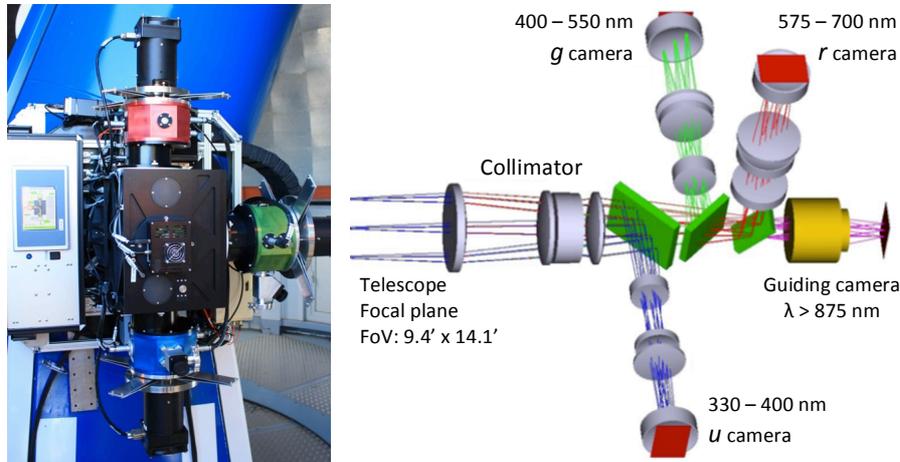

*Figure 2  Picture of MAIA mounted at the Nasmyth focus of the Mercator telescope (left) and optical layout of MAIA (right).*

Additionally, the presence of cryogenic temperatures in the cryostat that houses the detector, allows the use of a cryogenic sorption pump. A charcoal getter at a temperature of 77K is a simple method to obtain efficient adsorption of contaminating gasses in the cryostat (mostly water) and helps maintaining a good vacuum in the cryostat over extended periods of time. Condensation of contaminants mostly occurs on the surface with the lowest temperature in the cryostat. As a getter at cryogenic temperature will always be much colder than the detector, it offers some protection against contamination for the detector. Hence the availability of cryogenic temperatures (< 100K) in the cryostat is an important advantage.

### 3.2  Overview of cooling technologies

To cool the MAIA detectors to below 190K, we explored and compared various options. Because of its low cost, simplicity and high reliability, thermo-electric cooling with multi-stage Peltier elements was the method of first choice. Despite substantial engineering efforts, building prototypes with 3 and 4-stage Peltier stacks, it turned out that Peltier cooling for very large detectors like the CCD42-C0, was extremely difficult. Even with cooling water coming to aid, heat transfer from the cold detector to the hot side of the cooler remained limited. In combination with the very low efficiency of the stacked Peltier coolers, this resulted in minimum temperatures for the detector that were higher than 200K, well above the 190K requirement. Obviously, Peltier cooling excludes cryogenic temperatures, so cryo-pumping is not possible for maintaining a long-term vacuum in the cryostat.

The use of liquid nitrogen (LN$_2$), boiling off at a temperature of 77K, is a long-time proven and widespread solution to cool astronomical detectors. However, it requires the permanent presence of LN$_2$ at the observatory. This poses the strong operational burden of regularly refilling the cryostats with LN$_2$ (typically twice per day). The latter disadvantage is possibly avoided by a continuous-flow LN$_2$ system, but this cannot readily be implemented for a rotating instrument at the compact Nasmyth focal station of the Mercator telescope.

A second classic cooling solution in astronomy is a Joule-Thompson (JT) cooler like e.g. the CryoTiger (Brooks - Polycold, USA). This type of cooler also reaches LN$_2$ temperatures, provides plenty of cooling power and operates almost maintenance free. On the other hand, cooling the three MAIA cryostats implied preferably three compressors, dissipating 1500W in total, and six gas lines with heavy stainless steel braiding. Such a system is also too bulky to be accommodated at the Nasmyth focus of the Mercator telescope.

More mechanical cooling solutions exist, like e.g. Gifford-McMahon (GM) and pulse-tube (PT) cryogenic refrigerators, that both can reach much lower temperatures than required for MAIA. Nevertheless, they are less attractive for the MAIA application and for optical imaging in general. Very similar to JT, GM coolers also rely on a large external compressor. The performance of PT coolers depends on their orientation and they should preferably always be operated in vertical direction, meaning they are less useful for a rotating instrument.

The alternative that was finally selected to cool the MAIA detectors, is the use of compact free-piston Stirling coolers. These are heat engines that rely on the Stirling cycle to drive heat transfer. In this type of cryo-cooler, the piston is driven by a permanent magnet linear motor, meaning the device can be completely hermetically sealed and friction losses can be kept very small. As a consequence, these Stirling coolers have a long lifetime (> 10 years). The vibrations generated by the linear compressor are the only important drawback. In section 4 we discuss how these vibrations can be reduced to a level that is negligible for most astronomical applications.

### 3.3  Stirling coolers for the MAIA detectors

After some market research, the MT CryoTel Stirling cooler (SUNPOWER, USA) [3] was selected as cooling engine for the MAIA detectors. This cryo-cooler can evacuate 5W of heat at a temperature of 77K with an electrical input power of 80W. The performance of the cooler does not depend on its orientation, so it can be used conveniently in a rotating instrument or at a telescope's Cassegrain focus. It is an efficient and compact device with a weight of only 2.1 kg. The complete package, including cold finger, radiator, compressor body and vibration absorber, fits in the volume of a standard LN$_2$ bath cryostat. Figure 3 shows a picture of a CryoTel MT cooler with air-fin heat rejection, as is used for MAIA. Alternatively, heat can also be evacuated through a water jacket when local dissipation needs to be avoided.

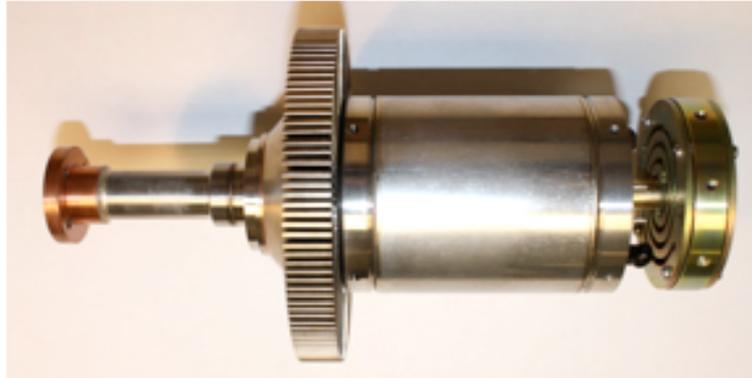
*Figure 3  CryoTel MT cryo-cooler.*

The piston of the CryoTel cooler is driven by a linear motor running at 60 Hz, causing vibrations at this frequency and its harmonics. Therefore, a resonating vibration absorber is mounted on the backside of the cooler. This resonator is precisely tuned to the cooler frequency and it absorbs the bulk of the vibration energy. Laboratory measurements show that this passive absorber effectively reduces vibrations at 60 Hz to less than 10% of their initial amplitude. However, the remaining vibrations can still degrade the performance of the instrument, meaning additional decoupling between the cooler and the cryostat or the instrument is required.

### 3.4    MAIA detector cooling system

The thermal link between the detector and the cold head of the cooler consists of 2 flexible copper braids (tinned grey braid in figure 1). Their length and section area are dimensioned to obtain a detector equilibrium temperature around 155K, with the cold head temperature set at 80K. This requires about 70W of electrical input power for the cooler. The MAIA cryostats are not equipped with a radiation shield that could limit the heat load on the cold head and the detector. Installing a radiation shield around the cold elements in the cryostat could potentially reduce the required cryo-cooler power.

A 100Ω resistive heater and a Pt100 temperature sensor are mounted at the back of the detector. They are used in closed loop to heat the detector (typical heating power is 2.4W) and stabilize its temperature at 165K with an accuracy of a few 0.01K. This feedback loop is implemented using a programmable logic controller (PLC) and off-the-shelf industrial hardware. This PLC also controls the Stirling cooler drivers and takes care of all other instrument control tasks (figure 4) [4]. The PLC is connected to a human-machine interface (HMI) with touch screen, mounted on the front side of the instrument. This HMI provides a visual indication of the status of the instrument (e.g. detector and cold head temperatures, actual cooling power, etc.) and allows low-level control of MAIA (e.g. changing temperature set-point for detector or cold head, setting control-loop parameters, starting a controlled warm-up procedure, etc.).

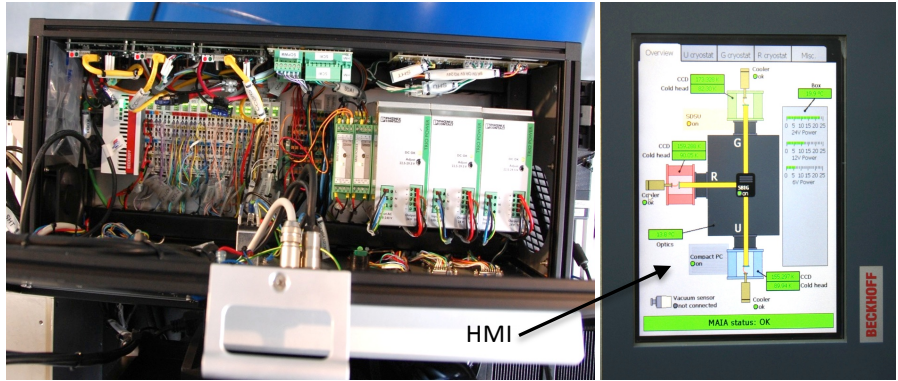

*Figure 4 left: MAIA electronics box (leftmost vertical part in figure 2) with PLC, power supplies and Stirling cooler controllers; right: HMI touch screen, mounted on door of box.*

## 4. VIBRATION CONTROL

### 4.1 Mechanical design

Initially, the cooler was rigidly fixed on the cryostat. Laboratory measurements showed that the cooler vibrations had no detrimental effect on image stability or image quality. However, after we installed MAIA on the Mercator telescope, significant vibrations of the telescope structure appeared. These vibrations induced unacceptable oscillations with amplitudes exceeding 5 arcsec. Spectral analysis of accelerometer measurements on the instrument and on the telescope showed that almost all of the vibration energy was contained within narrow-band signals at 60 Hz and its higher harmonic frequencies (figure 7, left). From the narrow-band nature of the telescope vibrations, we conclude that they are the result of direct coupling to the Stirling cooler, rather than the excitation of a resonant mode in the telescope structure. Intuitively, one may think that adding structural damping to the system is the best approach to attenuate these vibrations. However, this type of structural damping is more effectively used to attenuate wide-band signals [5]. Therefore, we decided to mechanically decouple the cooler from the rest of the instrument. We do this by mounting four leaf springs, consisting of 2-mm thick V-shaped aluminum blades, between the cooler and the cryostat. The vacuum interface between both parts consists of a short but flexible bellows. The cooler body is suspended in a plastic tube that also holds a fan to extract the heat from the cooler through the air-fin radiator. This is illustrated in figure 5, showing the layout of the cryostat and the mounting of the cooler.

The leaf springs create a mechanical low-pass filter that will attenuate the transmission of vibrations at frequencies that are well above the eigenfrequency of the system. The eigenfrequency or resonance frequency $f_{res}$ of a spring-mass system with $m$ the mass of the cryo-cooler and $k$ the total spring constant of the leaf springs and the vacuum bellows, is given by equation 1.

$$f_{res} = \frac{1}{2\pi}\sqrt{\frac{k}{m}} \qquad (1)$$

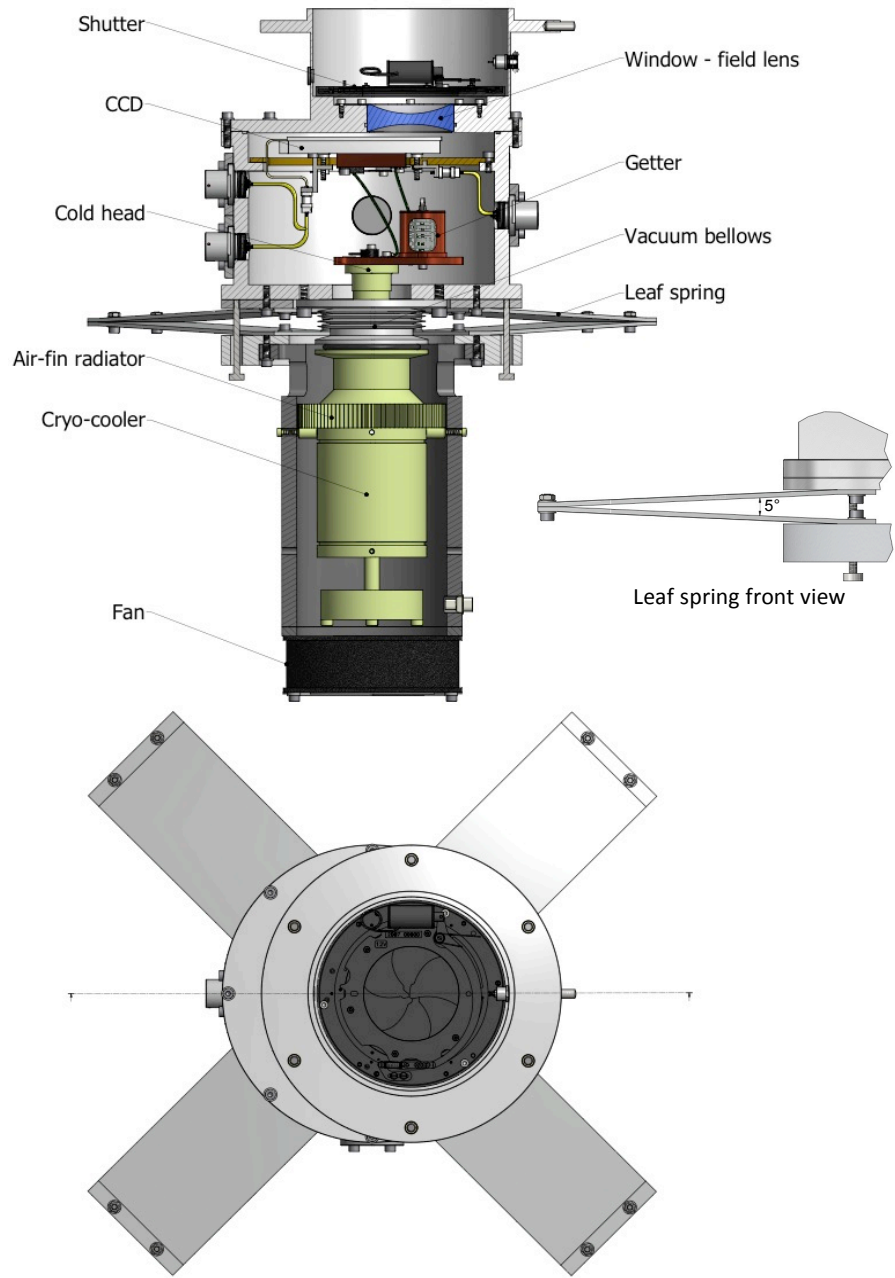

*Figure 5  Section and front view of the MAIA cryostat, detail of leaf spring.*

A low eigenfrequency (much smaller than the 60 Hz main frequency of the cooler) can be achieved by using springs with low stiffness. However, the springs should be sufficiently strong to resist the force exerted by the vacuum inside the cryostat (≈160 N at the altitude of the observatory), as well as to limit the axial and angular excursions of the cooler due to gravity when the instrument is rotated. These constraints lead to a lower limit for $k$ of 33500 N/m. We can reduce $f_{res}$ further by increasing $m$. Therefore we add extra mass to the cooler. Each cooler receives a stainless steel mounting ring of different weight (total cooler weight: 6.2, 7.2 and 8.5 kg for *r*, *u* and *g*). Different resonance frequencies for each cooler are the result (12, 11, and 10 Hz for *r*, *u* and *g*). This way, we avoid resonant coupling between the three spring-mass systems. Figure 6 (left) shows how forces from the vibrating Stirling cooler are transmitted through the springs to the cryostat.

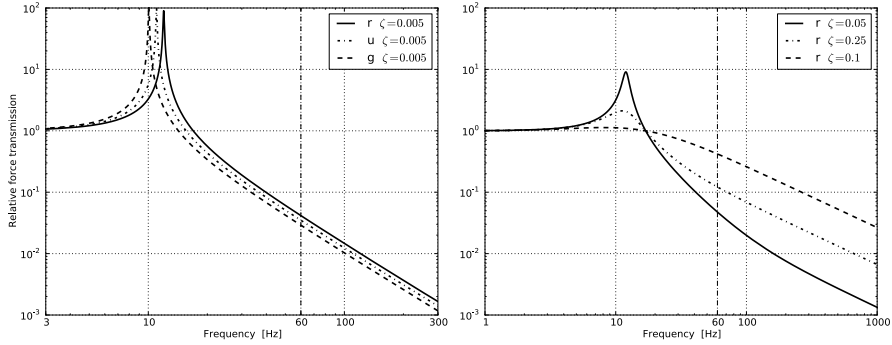

*Figure 6 Transmission of vibration forces through the spring-mass system from cooler to cryostat, left: for the three cryostats (u, g and r with resonance peaks at 11, 10 and 12 Hz respectively), without introducing any additional damping of the springs; right: for the red cryostat only, now with increasing damping ratios ζ to reduce the resonance peak gain.*

Besides $m$ and $k$, the force transmission at a given frequency is characterized by a third parameter, namely the damping ratio ζ, a unitless number that describes the oscillations' decay rate after a disturbance. Large damping results in a small resonance peak at the eigenfrequency but also in a reduced roll-off slope of the low-pass filter. This is illustrated in the right part of figure 6. It may seem counter-intuitive, but introducing damping between cooler and cryostat results in much less efficient attenuation at higher frequencies and should better be avoided. As we did not encounter any resonance problems, as little damping as possible (ζ≈0.005, a typical value for undamped aluminum blades) was added to the spring-mass system.

### 4.2 Performance of the vibration isolation system

To test the performance of the vibration isolation, the vibrational accelerations of both the instrument and the telescope were measured with and without the spring isolators installed. In figure 7, a typical vibration spectrum of the *r* cryostat is shown. Because *r* has the smallest additional mass and thus the highest eigenfrequency, this cryostat exposes the worst vibration case. We measured the accelerations on the cryostat, along the axis of the piston movement. Measurements at different locations and in different directions give very similar results but the amplitudes turn out smaller. The spring isolators reduce the harmonic peaks by several orders of magnitude. At the 60 Hz main frequency, this reduction amounts to a factor of 500. Although not really troublesome,

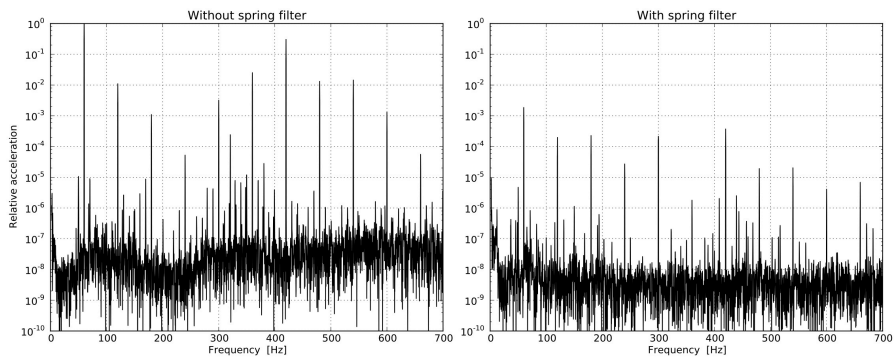

*Figure 7 Cryostat vibration spectrum, measured before (left) and after (right) installing the spring-mass decoupling between cooler and cryostat.*

broadband vibrational noise is also reduced by almost an order of magnitude at higher frequencies.

After we installed this vibration isolation system, extended series of images were taken during which we regularly switched 1, 2, or all coolers *On* and *Off*. Careful analysis of all these images could no longer reveal any effect on the image quality, nor on the position of the sources in the frames. We conclude that he effect of the remaining vibration level is smaller than the 0.1-arcsec detection limit of the image analysis that we performed.

Compared to many other facilities in astronomy, the MAIA instrument and the Mercator telescope are relatively small. On larger and more massive installations, any effect of cooler-induced vibrations is expected to be even smaller.

## 5. CONCLUSIONS

Compact free-piston Stirling coolers provide a convenient and attractive alternative to the traditional methods of cooling astronomical detectors. They are compact, power efficient, straightforward to use and can run maintenance-free for an extremely long time. Strong vibrations are the only important disadvantage, but that can be kept well under control by mounting the cooler with a spring system in a floating way on the cryostat. The MAIA imager successfully uses three Stirling coolers for cooling three CCDs in three separate cryostats. This application shows that through careful mechanical design, astronomical instruments can be built that do not suffer from any vibrations while using Stirling-cooled detectors. Although until now somewhat neglected in optical astronomical instruments, this type of detector coolers can be a useful option for many imaging applications with perhaps the exception of those with the most stringent stability requirements, like interferometers and high-accuracy radial-velocity spectrographs.

## 6. ACKNOWLEDGEMENTS

This research was based on funding from the European Research Council under the European Community's Seventh Framework Programme (FP7/2007–2013)/ERC grant agreement n°227224 (PROSPERITY) and from the Fund for Scientific Research of Flanders (FWO), grant agreements G.0410.09 and

## 7. REFERENCES

[1] G. Raskin, S. Bloemen, J. Morren, et al. "MAIA, a three-channel imager for asteroseismology: instrument design", Astronomy & Astrophysics, Vol. **559**, pp. A26, 2013.

[2] G. Raskin, G. Burki, M. Burnet, et al. "Mercator and the P7-photometer", SPIE Conference Series, Vol. **5492**, pp. 830, 2004.

[3] R. Unger and D. Keiter, "The development of the CryoTel family of coolers", AIP Conference Proceedings, Vol. **710**, pp. 1404, 2004.

[4] W. Pessemier, G. Deconinck, G. Raskin, et al. "Design and first commissioning results of PLC-based control systems for the Mercator telescope", SPIE Conference Series, Vol. **8451**, pp. 84512V, 2012.

[5] J. Den Hartog, *Mechanical Vibrations* (Dover Publications – Dover Books on Engineering, 1985).